\documentclass{article}

\title{Dirac-Coulomb scattering with plane wave energy eigenspinors on de Sitter expanding universe}
\author{Ion I.  Cot\u aescu and Cosmin Crucean\\
{\normalsize West University of Timi\c soara,  V. Parvan}\\
\normalsize Avenue 4 RO-300223 Timi\c soara,  Romania}
\begin{document}
\maketitle

\begin{abstract}
The lowest order contribution of the amplitude of Dirac - Coulomb scattering in
de Sitter spacetime is calculated assuming that the initial and final states of
the Dirac field are described by exact solutions of the free Dirac equation on
de Sitter spacetime with a given energy and helicity.  We find that the total
energy is conserved in the scattering process.
\end{abstract}

\section{Introduction}
\par
Recently a new time-evolution picture of the Dirac quantum mechanics was
defined in charts with spatially flat Robertson-Walker metrics,  under the name
of Scr\" odinger picture \cite{B1}.  Using the advantage offered by this
picture in \cite{B2} was found a new set of Dirac energy eigenspinors which
behave as polarized plane waves.
\par
Also recently the Coulomb scattering on de Sitter expanding universe was
studied \cite{B4},  using the plane wave solutions derived in \cite{B3}.  The
main results in \cite{B4} was that the modulus of total momentum is not
conserved but there is a tendency of helicity conservation.
\par
On the other hand, the expansion of the Universe is accelerating,  and this
could increase the interest of studying the scattering processes on de Sitter
backgrounds.  In the present paper we would like to analyse the Coulomb
scattering using the method of \cite{B4}  with the energy eigenspinors derived
in \cite{B2}, pointing out a series of aspects in comparison with the study
presented in \cite{B4}.  We shall see that in this case  the scattering has new
important features since this time the the energy is conserved.
\par
The paper is organized as follows.  In section 2,  we present a short review of
the Schr\" odinger picture introduced in \cite{B1} and we write the form of the
energy eigenspinors derived in \cite{B2}.  In Section 3 we define the lowest
order contribution for scattering amplitude in the potential $A^{\hat{\mu}}$ in
the new Schr\" odinger picture and then we calculate the scattering amplitude,
showing that the total energy conservation holds in this case. Section 4 is
dedicated to the problem of cross section.  Our conclusions are summarized in
section 5 pointing out a series of aspects that remain to be clarified
elsewhere.
\par
We use elsewhere natural units i. e.  $\hbar=c=1$.

\section{Polarized plane wave in the Schr\" odinger picture}

We start with the results of \cite{B1},  where was shown that two time
evolution pictures can be identified in the case of the Dirac theory on
backgrounds with spatially flat Robertson-Walker metrics.  The idea there was
to define the natural picture as that in which the free Dirac equation is
written directly as it results from its Lagrangean,  in a diagonal gauge and
Cartesian coordinates. In addition the Schr\" odinger picture was introduced in
which the free Dirac equation is transformed such that its kinetic part takes
the same form as in special relativity while the gravitational interaction is
separated in a specific term.
\par
Let us take the local chart with cartesian coordinates of a flat
Robertson-Walker manifold,  in which the line element reads:
\begin{equation}\label{1}
ds^{2}=dt^{2}-\alpha(t)^{2}d\vec{x}^{2},
\end{equation}
where $\alpha$ is an arbitrary function. One knows that for defining  spinor
fields on curved backgrounds it requires to introduce the tetrad fields
$e_{\hat{\mu}}(x)$ and $\hat{e}^{\hat{\mu}}(x)$,  fixing the local frames and
corresponding coframes which are labelled by the local indices $\hat{\mu},
\hat{\nu}=0, 1, 2, 3$.  The form of the line element allows one to choose the
simple diagonal gauge where the tetrad fields have the non-vanishing
components\cite{B2}, \cite{B3}:
\begin{equation}\label{2}
e^{0}_{\hat{0}}=1 \quad , e^{i}_{\hat{j}}=\frac{1}{\alpha(t)}\delta^{i}_{j},
\quad \hat{e}^{0}_{0}=1, \quad \hat{e}^{i}_{j}=\alpha(t)\delta^{i}_{j}.
\end{equation}
The Dirac field $\psi$ of mass $m$ satisfy the free Dirac equation which can be
easily written using the tetrad fields (\ref{2}) (see \cite{B2}).  If $\psi(x)$
is the Dirac field in natural picture then the Dirac field of the Schr\"
odinger picture,$\psi_{S}(x)$, can be obtained using the transformation
$\psi_{S}(x)=W(x)\psi(x)$  produced by the operator of time dependent
dilatations \cite{B1},
\begin{equation}\label{3}
W(x)=exp\left[-\ln(\alpha(t))(\vec{x}\cdot\vec{\partial})\right],
\end{equation}
which has the property
\begin{equation}\label{4}
W(x)^{+}=\sqrt{-g(t)}W(x)^{-1}\,.
\end{equation}
Using this operator, the Dirac equation of the Schr\" odinger
picture was obtained in \cite{B2}, as well as the relativistic
scalar product $\langle\psi_{S}, \psi^{'}_{S}\rangle=\int d^{3}x
\bar{\psi}_{S}(x)\gamma^{0}\psi^{'}_{S}(x)$, which is no more
dependent of $\sqrt{-g(t)}$.
\par
Now taking in Eq. (\ref{1}) $\alpha(t)=e^{\omega t}$ one obtains the de Sitter
metric which is the case of interest here.  The form of the Dirac equation on
de Sitter spacetime in Schr\" odinger picture is given in \cite{B2} where a
complete set of orthonormalized funndamental solutions was written down. These
depend on  the normalized Pauli spinors, $\xi_{\lambda}(\vec{n})$, of {\em
helicity} $\lambda=\pm 1/2$ which satisfy
\begin{equation}\label{5}
(\vec{n}\cdot\vec{\sigma})\xi_{\lambda}(\vec{n})=2 \lambda
\xi_{\lambda}(\vec{n}),
\end{equation}
where $\vec{\sigma}$ are the Pauli matrices while the momentum direction is
given by $\vec{n}$ ($\vec{p}=p\vec{n}$). Then the fundamental spinor solutions
of positive frequencies with energy $E$, momentum direction $\vec{n}$ and
helicity $\lambda$ obtained in \cite{B2} read
\begin{equation}\label{6}
U^{S}_{E, \vec{n}, \lambda}(t, \vec{x})=i\frac{ \omega
e^{-iEt}}{(2\pi)^{3/2}\sqrt{2}}\int^{\infty}_{0}s ds\left (\begin{array}{c}
\frac{1}{2}e^{\pi k/2}H^{(1)}_{\nu_{-}}(s)\xi_{\lambda}(\vec{n})\\
\lambda e^{-\pi k/2}H^{(1)}_{\nu_{+}}(s)\xi_{\lambda}(\vec{n})
\end{array}\right)e^{i\omega s\vec{n}\vec{x}-i\epsilon \ln s}\,.
\end{equation}
The notations used here are $\nu_{\pm}=\frac{1}{2}\pm ik$ with $k=m/\omega$,
$s=p/\omega$ and $\epsilon=E/\omega$.  The negative frequency modes can be
obtained using the charge conjugation as in \cite{B2} $U^{S}_{E, \vec{n},
\lambda}(x)\rightarrow V^{S}_{E, \vec{n},
\lambda}(x)=i\gamma^{2}\gamma^{0}(\bar{U}^{S}_{E, \vec{n}, \lambda}(x))^{T}$,
because the charge conjugation in a curved space is point independent
\cite{B5}. However the negative frequency modes will be of no interest here.

These spinors are normalized in the energy scale (in generalized sense) with
respect to the new relativistic scalar product defined in Scr\" odinger picture
\cite{B2}:
\begin{eqnarray}
&&\int d^{3}x \bar{U}^{S}_{E, \vec{n}, \lambda}\gamma^{0}
U^{S}_{E^{'}, \vec{n}^{'}, \lambda^{'}}=\nonumber\\
&&~~~~~~~~~~\int d^{3}x \bar{V}^{S}_{E, \vec{n}, \lambda}\gamma^{0}
V^{S}_{E^{'}, \vec{n}^{'},
\lambda^{'}}=\delta_{\lambda\lambda^{'}}\delta(E-E^{'})
\delta^{2}(\vec{n}-\vec{n}^{'}),\label{7}
\end{eqnarray}
where $\delta^{2}(\vec{n}-\vec{n}^{'})=\delta(\cos \theta_{n}-\cos
\theta^{'}_{n})\delta(\phi_{n}-\phi^{'}_{n})$.  These spinors form a complete
system of solutions:
\begin{eqnarray}
&&\int_{0}^{\infty} dE \int_{S^{2}}d\Omega_{n}
\sum_{\lambda}\left[U_{E, \vec{n}, \lambda}(t, \vec{x})U^{+}_{E, \vec{n},
 \lambda}(t, \vec{x^{\prime}})\right.\nonumber\\
&&\hspace*{30mm}\left.+V_{E, \vec{n}, \lambda}(t, \vec{x})V^{+}_{E, \vec{n},
\lambda}(t, \vec{x^{\prime}})\right]=
\delta^{3}(\vec{x}-\vec{x^{\prime}}).\label{8}
\end{eqnarray}

\section{The scattering amplitude}

 The solutions written in \cite{B2} will be the central piece of our
 calculations.  In \cite{B4} it was pointed out that the necessary requirement
 for developing the scattering on de Sitter background is the global
  hyperbolicity of the spacetime and having a complete set of solutions of the free
  equation for incident and scattered field (Born approximation).
  Now for defining the lowest order contribution of the scattering amplitude in the
 Schr\" odinger picture let us recall the definition of this quantity
 from \cite{B4} in the natural picture:
 \begin{equation}\label{9}
 A_{i\rightarrow f}=-ie \int d^{4}x
\left[-g(x)\right]^{1/2}\bar\psi_{f}(x)\gamma_{\mu}A^{\hat{\mu}}(x)\psi_{i}(x).
 \end{equation}
 This expression was obtained by analogy with Minkowski space \cite{B6,B9},
 but can also be obtained from one reduction formalism on de
 Sitter spacetime \cite{B8}.  Using now (\ref{4}) it is not hard to obtain the
 analogue of (\ref{8}) in the Schr\" odinger picture:
\begin{equation}\label{10}
 A^{S}_{i\rightarrow f}=-ie \int d^{4}x
\bar\psi_{Sf}(x)\gamma_{\mu}A^{\hat{\mu}}_{S}(x)\psi_{Si}(x),
 \end{equation}
 where $e$ is the unit charge of the field,   $A^{\hat{\mu}}_{S}(x)$ is the potential in the
 Schr\" odinger picture,  and the hated indices label the
 components in local Minkowski frames.
 \par
 Our target is a fixed charge $Ze$ whose  Coulomb potential on de Sitter
 spacetime in the natural picture \cite{B4} reads
 \begin{equation}\label{11}
A^{\hat{0}}(x)=\frac{Ze}{|\vec{x}|} e^{-\omega t},
\end{equation}
while in the new Schr\" odinger picture this becomes
\begin{equation}\label{12}
A^{\hat{0}}_{S}(x)=\frac{Ze}{|\vec{x}|}\,.
\end{equation}
\par
Our aim  is to calculate the amplitude of Coulomb scattering using
the definition (\ref{10}) in which we replace our quantities of
interest (\ref{6}) and (\ref{12}).  We start with the waves freely
propagating in the $in$ and $out$ sectors, $U^{S}_{E_{i}, \vec{n},
\lambda_{i}}(x)$ and $U^{S}_{E_{f}, \vec{n}, \lambda_{f}}(x)$,
assuming that the both of them are of positive frequency.  If we
replace the explicit form of spinors and the Coulomb potential in
(\ref{10}) we observe two remarkable properties.  The first one is
that we may split the four dimensional integral into a pure
spatial integral and a temporal one. In other respects, these
integrals have the same form as in Minkowski spacetime, i. e.
\begin{eqnarray}
\int d^{3}x \frac{e^{i(\vec{p_{i}}-\vec{p_{f}})\vec{x}}}{|\vec{x}|}&=&\frac{4
\pi}{|\vec{p_{f}}-\vec{p_{i}}|^{2}}\,, \nonumber\\
\frac{1}{2 \pi}\int^{\infty}_{-\infty} dt
e^{i(E_{f}-E_{i})t}&=&\delta(E_{f}-E_{i})\,.\label{13}
\end{eqnarray}
Note that the limits of integration in (\ref{13}) for the time
variable correspond to $t=\pm \infty$,  assuming that the
interaction extends into the past and future.
\par
In this case the integration after $s=p/\omega$ variable is not quite simple
but we can calculate our amplitude as
\begin{eqnarray}
&& A^{S}_{i\rightarrow f}= \frac{-i\alpha Z
\omega^{2}\delta(E_{f}-E_{i})}{8\pi|\vec{p_{f}}-\vec{p_{i}}|^{2}}\xi^{+}_{\lambda_{f}}
\,(\vec{n})\xi_{\lambda_{i}}(\vec{n})\nonumber\\
&&\times\left[e^{\pi k}\int_0^{\infty} ds_{f}
s^{1+iE_{f}/\omega}_{f}H^{(2)}_{\nu_{+}}(s_{f})\int_{0}^{\infty}
ds_{i}s^{1-iE_{i}/\omega}_{i}
H^{(1)}_{\nu_{-}}(s_{i})\right. \nonumber\\
&&\left. +sgn(\lambda_{f}\lambda_{i})e^{-\pi k}\int_0^{\infty}
ds_{i}s^{1+iE_{f}/\omega}_{f}H^{(2)}_{\nu_{-}}(s_{f})
\int_{0}^{\infty}
ds_{i}s^{1-iE_{i}/\omega}_{i}H^{(1)}_{\nu_{+}}(s_{i})\right]\label{14}
\end{eqnarray}
where $\alpha=e^{2}$ . The evaluation of the integrals (\ref{14})
is given in Appendix A, the final result being expressed in terms
of Euler gamma functions,
\begin{eqnarray}
A^{S}_{i\rightarrow f}&=& \frac{-i\alpha Z
\omega^{2}\delta(E_{f}-E_{i})}{4\pi|\vec{p}_{f}-\vec{p}_{i}|^{2}}\,
\xi^{+}_{\lambda_{f}}(\vec{n})\xi_{\lambda_{i}}(\vec{n})\nonumber\\
&&\times\left[
f_{k}(E_{f})f^{*}_{k}(E_{i})+sgn(\lambda_{f}\lambda_{i})
f_{-k}(E_{f})f^{*}_{-k}(E_{i})\right]\label{15}
\end{eqnarray}
where we introduced the following notations:
\begin{eqnarray}
&& f_{k}(E)=e^{\pi
k/2}\left[2^{iE/\omega}\frac{\Gamma(\frac{5}{4}+\frac{i
k}{2}+\frac{i E}{2 \omega})}{\Gamma(\frac{1}{4}+\frac{i
k}{2}-\frac{i E}{2
\omega})}-\frac{i2^{iE/\omega}}{\pi}\cos\left(\frac{\pi}{4}+\frac{i
k
\pi}{2}-\frac{i E \pi}{2 \omega}\right)\right.\nonumber\\
&&\left.\times \Gamma\left(\frac{5}{4}+\frac{i k}{2}+\frac{i E}{2
\omega}\right)\Gamma\left(\frac{3}{4}-\frac{i k}{2}+\frac{i E}{2
\omega}\right)\right],\label{16}
\end{eqnarray}
and $f_{-k}(E)$ is obtained when $k \rightarrow -k$ in (\ref{16}).

Now let us take a look to our scattering amplitude (\ref{15}).  We
obtain that the energy is conserved in the scattering process as
in Minkowski case.  This is however expected because the form of
the external field (\ref{12}) allows us to consider that the
scattering process take place in a constant field.  One knows that
the energy of a system scattered on a constant field  is conserved
(but this does not mean that the momentum is conserved too), as we
obtained here. It is also remarkable that we obtain the Rutherford
denominator $|\vec{p_{f}}-\vec{p_{i}}|^{2}$ as in Minkowski
scattering.  In our previous work \cite{B4} the Coulomb scattering
was analyzed with spinors having given momentum and helicity.  The
surprising result was that there exists a nonvanishing probability
for a scattering process where the law of conservation of total
momentum is lost.  Here we obtain the nice result that the total
energy is always conserved and the non-linear terms that may
broken the energy conservation have no contributions to the
amplitude (\ref{15}).
\par
Let us make an analysis in the helicity space.   We observe that
the analysis can be done using only the terms from parenthesis in
(\ref{15}).  Now the probability of scattering is proportional
with the amplitude at square $P \sim |A_{i\rightarrow f}|^{2}$.
After a little calculation we obtain that the probability of
transition between identical helicity states is bigger that the
probability of transition between opposite helicity states
($P_{\lambda_{i}=\lambda_{f}}>P_{\lambda_{i}\neq\lambda_{f}}$)
with the quantity:
\begin{equation}\label{17}
2\left[f_{k}(E_{f})f^{*}_{-k}(E_{f}) f^{*}_{k}(E_{i})f_{-k}(E_{i})
+f^{*}_{k}(E_{f})f_{-k}(E_{f}) f^{*}_{-k}(E_{i})f_{k}(E_{i})\right]
\end{equation}
Hereby we conclude that in the scattering process  a tendency is manifested for
helicity conservation.  This conclusion was also obtained in \cite{B4} were the
analysis was done using the momentum eigenspinors. The obvious conclusion is
that in the de Sitter space there is a tendency for total angular momentum
conservation.

\section{The cross section problem}

The first observation here is that in this case we have just linear
contributions to the cross section in contrast with \cite{B4}  where the cross
section was calculated as a sum of a linear contribution and a non-linear one.
Moreover, the term $\delta(E_{f}-E_{i})$ will give us the opportunity to define
the transition probability in unit of time like in Minkowski case.  Then the
definition of cross section here will have the same form as in Minkowski space,
\begin{equation}\label{18}
d\sigma=\frac{1}{2}\sum_{\lambda_{i}\lambda_{f}}\frac{dP}{dt}\frac{1}{j},
\end{equation}
where $\frac{dP}{dt}$ is the transition probability in unit of time, $j$ is the
incident flux while the factor $\frac{1}{2}$ transforms the sum in mediation.
\par
The problem of calculating the incident flux is identic to that of \cite{B4}.
First let us introduce the expression of the Dirac current in local frames,
\begin{equation}\label{19}
J^{\hat\mu}=e^{\hat\mu}_{\nu}\bar U_{\vec{p_{i}},
\lambda_{i}}(x)\gamma^{\nu}U_{\vec{p_{i}}, \lambda_{i}}(x)\,.
\end{equation}
Then the spatial components can be defined as follows:
\begin{eqnarray}
 j(t)&=&e^{\omega
t}\bar{U}^{S}_{E_{i}, \vec{n}, \lambda_{i}} (x)(\vec{n}\cdot\vec{\gamma}
)U^{S}_{E_{i}, \vec{n}, \lambda_{i}}(x)\nonumber\\
&=&\frac{e^{\omega t}\omega^{2}}{32
\pi^{3}}\left|\left[1+i\cot\left(\frac{\pi}{4}+\frac{ik\pi}{2}\right)\right]\left(\frac{1}{2}
+ik\right)\right|^{2}\,.\label{20}
\end{eqnarray}
From this equation it is immediate that our incident flux is a time dependent
quantity. We know from the well-establish picture from Minkowski spacetime that
the incident flux does not depend of time.  This property is no longer valid in
a spacetime where the translational invariance with respect to time is lost,
and our result is in agreement with this observation.  We note that in
\cite{B4} it was obtained a similar dependence of time for the incident flux.
Another observation is that the incident flux calculated here does not depend
on the incident momentum like in \cite{B4}.
\par
The cross section however must be evaluated using an incident flux that is
independent on time.  We will follow the formalism presented in \cite{B7},
where for the calculation of the incident flux one must know the state of the
unperturbed system from the approximative moment of collision.  In \cite{B7},
this was taken to be $t\sim 0$, which in the case of our incident flux
(\ref{20}) gives:
\begin{equation}\label{21}
j=j(0)=\frac{\omega^{2}}{32 \pi^{3}}\frac{2 e^{\pi k}}{\cosh(\pi
k)}\left(k^2+\frac{1}{4}\right)\,.
\end{equation}
\par
The evaluation of the transition probability in unit of time
$\frac{dP_{l}}{dt}=\frac{d|A^{S}_{i\rightarrow
f}|^{2}}{dt}\frac{d^{3}p_{f}}{(2\pi)^{3}}$ ( where we use the fact that
$[\delta(E_{f}-E_{i})]^{2}=\frac{t}{2\pi}\delta(E_{f}-E_{i})$) yields
\begin{eqnarray}
&& \frac{dP_{l}}{dt}=\frac{(\alpha Z)^{2}\omega^{4}}{32
\pi^{3}|\vec{p}_{f}-\vec{p}_{i}|^{4}}\delta(E_{f}-E_{i})\left[
|f_{k}(E_{f})|^{2}|f_{k}(E_{i})|^{2}\right. \nonumber\\
&&\left. +|f_{-k}(E_{f})|^{2}|f_{-k}(E_{i})|^{2}
+sgn(\lambda_{f}\lambda_{i})f_{k}(E_{f})f^{*}_{-k}(E_{f})
f^{*}_{k}(E_{i})f_{-k}(E_{i})\right. \nonumber\\
&&\left. +sgn(\lambda_{f}\lambda_{i})f^{*}_{k}(E_{f})f_{-k}(E_{f})
f^{*}_{-k}(E_{i})f_{k}(E_{i})\right]
[\xi^{+}_{\lambda_{f}}(\vec{n})\xi_{\lambda_{i}}(\vec{n})]^{2}
\frac{d^{3}p_{f}}{(2\pi)^{3}}\,.\label{22}
\end{eqnarray}
\par
For obtaining the cross section when we have particles with given helicities we
must average upon the helicities of incident particles and sum upon the
helicities of emergent ones.  In our case we obtain :
\begin{equation}\label{23}
\frac{1}{2}\sum_{\lambda_{i}\lambda_{f}}
\left[\xi^{+}_{\lambda_{f}}(\vec{n})\xi_{\lambda_{i}}(\vec{n})\right]^{2}=2\,.
\end{equation}
\par
The final expression of the differential cross section after we replace
(\ref{21}), (\ref{22}) and (\ref{23}) in (\ref{18}) turns out to be:
\begin{eqnarray}
&&
 d\sigma=\frac{(\alpha Z)^{2}\omega^{2}}{4
\pi^{3}|\vec{p}_{f}-\vec{p}_{i}|^{4}}\frac{1+e^{-2\pi k}}{1+4
k^2}\,\delta(E_{f}-E_{i})
 \left[
|f_{k}(E_{f})|^{2}|f_{k}(E_{i})|^{2}\right. \nonumber\\
&&\left. +|f_{-k}(E_{f})|^{2}|f_{-k}(E_{i})|^{2}
+sgn(\lambda_{f}\lambda_{i})f_{k}(E_{f})f^{*}_{-k}(E_{f})
f^{*}_{k}(E_{i})f_{-k}(E_{i})\right. \nonumber\\
&&\left. +sgn(\lambda_{f}\lambda_{i})f^{*}_{k}(E_{f})f_{-k}(E_{f})
f^{*}_{-k}(E_{i})f_{k}(E_{i})\right]d^{3}p_{f}\,.\label{24}
\end{eqnarray}
\par
In Minkowski case the factor with $\delta(E_{f}-E_{i})$ was eliminated after
performing the integral with respect to the final momentum,  because there the
relation between momentum and energy is known.  In de Sitter spacetime we do
not know this relation, and the factor that contain delta Dirac distribution
can not be eliminated when one performs the integration with respect to final
momentum in (\ref{24}).
\par
We note that in \cite{B4} integrals of this type were used for
evaluating the cross section. These had the form
\begin{eqnarray}
\int_{0}^{\infty}dp_{f}
f(p_{f})p^{2}_{f}\delta(p_{f}-p_{i}), \nonumber\\
\int_{0}^{\infty}dp_{f} f(p_{f})p^{2}_{f}\theta(p_{f}-p_{i})\,,\label{25}
\end{eqnarray}
since  there we calculated the scattering process using spinors
with a definite momentum but unknown energy. The last integral in
(\ref{25}) was discussed in \cite{B4} ($\theta(p_{f}-p_{i})$ is
the unit step function),  and is solved when the modulus of the
momentum is not conserved in the scattering process.
\par
 We observe that our cross sections have a complicated dependence of
 energy which is quite unusual.  This dependence of energy
was obtained after the integration with respect to $s=p/\omega$,
which means that this dependence translated in physical terms
means that our cross section is still dependent of the form of the
incident wave.  However one can write
$d^{3}p_{f}=p^{2}_{f}dp_{f}d\Omega_{p_{f}}$ and solve the integral
with respect to the final momentum in (\ref{24}) for obtaining
$\frac{d\sigma}{d\Omega}$ but restricting the limits of
integration between zero and a maximal value of the final
momentum.
\par
 Finally it will be interesting to
study how the results obtained here in the Schr\" odinger picture have to be
translated in the natural picture.  This is because in the natural picture the
potential is no longer constant. First of all one knows from \cite{B2} that the
fundamental spinor solution of positive frequency (\ref{6}) can be wreathed in
the natural picture as $U_{E_{i}, \vec{n},
\lambda_{i}}(t,\vec{x})=U^{S}_{E_{i}, \vec{n}, \lambda_{i}}(t,\vec{x}e^{\omega
t})$. In addition, the external Coulomb field  in the natural picture is given
by Eq. (\ref{11}). Replacing these quantities in the definition of the
scattering amplitude from natural picture (\ref{9}), one will obtain the same
scattering amplitude as (\ref{15}). This can be checked out passing to a new
variable of integration $y=xe^{\omega t}$ when one solves the spatial
integrals. It means that our main conclusions from this paper (energy
conservation and the tendency for helicity conservation) will also remain valid
in the natural picture.
\section{Conclusion}
\par
We examined in this paper the Coulomb scattering on de Sitter spacetime using
the energy eigenspinors.  In our considerations the initial and final states of
the field are described by exact solutions (with a given energy and helicity)
of the free Dirac equation on de Sitter space,  which were written in the
Schr\" odinger picture.
\par
Moreover, we found that the scattering amplitude and the cross sections depend
on the expansion factor as $\omega^{2}$.  In addition we recover the result
from our previous work that the amplitude and implicitly the cross section
depends on the form of the incident wave.  The incident flux was also found as
a time dependent quantity.  Needless to say that this consequences is the
result of the lost translational invariance with respect to time in de Sitter
spacetime.
\par
 In section 3 we found that the total energy is
conserved in the scattering process and, in addition, terms that
could broken the energy conservation are absent since the
scattering was considered in a constant field of the form
(\ref{12}).  In section 3 we recover the tendency for helicity
conservation as in \cite{B4}.
\par
For further investigations it will be interesting to obtain the
definition of the scattering amplitude (\ref{10}) in the new
Schr\" odinger picture from one reduction formalism for the Dirac
field. This will require to use the form of the Dirac equation in
the Schr\" odinger picture \cite{B2} and the fundamental spinor
solutions (\ref{6}), with the distinction between
positive/negative frequencies.
\section{Appendix A}

The integrals that help us to arrive at the scattering amplitude
(\ref{15}) are of the type:
\begin{eqnarray}
&&\int_0^{\infty} dz
z^{1-iE/\omega}H^{(1)}_{\mu}(z)=2^{1-iE/\omega}\frac{\Gamma(\frac{\mu}{2}+1-\frac{i
E}{2\omega})} {\Gamma(\frac{\mu}{2}+\frac{i
E}{2\omega})}\nonumber\\
&&+i\frac{2^{1-iE/\omega}}{\pi}\cos\left(\frac{\mu \pi}{2}+\frac{i E
\pi}{2\omega}\right)\Gamma\left(-\frac{\mu}{2}+1-\frac{i
E}{2\omega}\right)\Gamma\left(\frac{\mu}{2}+1-\frac{i
E}{2\omega}\right)\label{26}
\end{eqnarray}
and
\begin{eqnarray}
&&\int_0^{\infty} dz
z^{1+iE/\omega}H^{(2)}_{\mu}(z)=2^{1+iE/\omega}\frac{\Gamma(\frac{\mu}{2}+1+\frac{i
E}{2\omega})} {\Gamma(\frac{\mu}{2}-\frac{i
E}{2\omega})}\nonumber\\
&&-i\frac{2^{1+iE/\omega}}{\pi}\cos\left(\frac{\mu \pi}{2}-\frac{i E
\pi}{2\omega}\right)\Gamma\left(-\frac{\mu}{2}+1+\frac{i
E}{2\omega}\right)\Gamma\left(\frac{\mu}{2}+1+\frac{i
E}{2\omega}\right)\,.\label{27}
\end{eqnarray}
Now setting $z=p/\omega$ and $\mu=1/2\pm ik$ one can see that our
result (\ref{15}) is correct.
\par
For calculating our incident flux we solve integrals of the form:
\begin{eqnarray}
\int_0^{\infty} dz z H^{(1)}_{\mu}(z)=\mu+i\mu\cot\left(\frac{\mu
\pi}{2}\right)\,, \nonumber\\
\int_0^{\infty} dz z H^{(2)}_{\mu}(z)=\mu-i\mu\cot\left(\frac{\mu
\pi}{2}\right)\,.\label{28}
\end{eqnarray}


\begin{thebibliography}{99}
\bibitem{B1}
 I. I. Cot\u aescu, {\em Mod. Phys. Lett.  A} accepted  (2007)
\bibitem{B2}
 I. I. Cot\u aescu and  C. Crucean,  qr-qc/0711. 0816, (2007)
\bibitem{B4}
 C. Crucean, {\em Mod. Phys. Lett.  A}  \textbf{22},  (2007)
\bibitem{B3}
 I. I. Cot\u aescu, {\em Phys. Rev.  D } \textbf{65}, (2002)
\bibitem{B5}
I. I. Cot\u aescu, {\em  Mod. Phys. Lett.  A}  \textbf{19},  (2004)
\bibitem{B6}
 L. Landau and E. M. Lifsit,  {\em Theorie Quantique Relativiste}
(Mir Moscou 1972)
\bibitem{B9}
  S. Drell and J. D. Bjorken,  {\em Relativistic Quantum Fields}
(Me Graw-Hill Book Co. ,  New York 1965)
\bibitem{B8}
  C. Crucean and  R. Racoceanu,  hep-th/0707. 1200,  (2007)
\bibitem{B7}
 M. L. Goldberger and  K. M. Watson,  {\em Collision theory } (Wiley,  New York,  1964)
\end{thebibliography}
\end{document}